# Sub-10 nm Nanochannels Enable Directional Quasi-Ballistic Exciton Transport over 5 μm at Room Temperature

Xiao-Jie Wang[1†], Jia-Wei Tan[3†], Xiao-Ze Li[1†], Hong-Hua Fang[1*], Guan-Yao Huang[1], Yang-Yi Chen[1], Yuan Luo[2], Jia-Tai Huang[1], Gong Wang[3], Qi-Hua Xiong[2], Xavier Marie[4,5], Hong-Bo Sun[1*]

[1]State Key Laboratory of Precision Measurement Technology and Instruments, Department of Precision Instrument, Tsinghua University, Beijing, 100084 China

[2]State Key Laboratory of Low-Dimensional Quantum Physics, Department of Physics, Tsinghua University, Beijing 100084, China

[3]Center for Advanced Laser Technology, Hebei University of Technology, Tianjin 300401, China

[4]Université de Toulouse, INSA-CNRS-UPS, LPCNO, 135 Avenue Rangueil, 31077, Toulouse, France

[5]Institut Universitaire de France, 75231, Paris, France

*Corresponding author:

Hong-Hua Fang, *hfang@mail.tsinghua.edu.cn*

Hong-Bo Sun, *hbsun@tsinghua.edu.cn*

†These authors contributed to this work equally.

# ABSTRACT

Nanoscale potential wells provide a powerful means to engineer energy landscapes in low-dimensional materials, enabling control over quantum states, carrier dynamics, and optoelectronic responses. Such confinement governs phenomena including charge localization, transport anisotropy, band structure modulation, and light–matter interaction strength. However, realizing clean and well-defined nanostructures remains technically challenging, as fabrication techniques such as focused ion beam (FIB) milling and electron-beam lithography frequently introduce structural disorder, residual contamination, or detrimental interactions with the underlying substrate. Here, we develop a femtosecond laser direct-writing technique to create sub-10 nm-wide dielectric nanochannels with smooth, continuous boundaries on hexagonal boron nitride (hBN) substrates, without using resists or chemical etchants. As a demonstration, these nanochannels are employed to define programmable dielectric landscapes in monolayer molybdenum diselenide ($MoSe_2$), forming excitonic energy funnels that suppress scattering and significantly extend the exciton transport distance. Transport is reshaped from isotropic diffusion with sub-micron range to directional super-diffusion exhibiting quasi-ballistic transport exceeding 5 μm, more than 20 times longer than in unpatterned systems. The smooth dielectric boundaries further enable precise control over exciton trajectories, allowing for programmable transport pathways. This dry, scalable, and substrate-compatible approach offers a robust platform for exciton engineering and integrated quantum photonic devices.

**INTRODUCTION**

Nanoscale potential wells are fundamental building blocks for controlling quantum states and tailoring electronic, optical, and excitonic properties in low-dimensional semiconductors[1-4]. At these length scales, local energy modulation can induce carrier confinement, band structure reshaping, enhanced light–matter interactions, and quantum-level control that are not achievable in bulk or uniformly structured systems[5-7]. Such control underpins a wide range of applications, including quantum light sources, photodetectors, nonlinear optical elements, and excitonic circuits[8-12]. A representative example is exciton transport in monolayer materials, where excitons are highly sensitive to their surrounding potential landscape[13-16]. In the absence of confinement, they exhibit isotropic diffusion with limited propagation length. By introducing well-defined nanoscale potential wells or guiding structures[17-21], exciton flow can be directed over long distances (Fig. 1a), enabling directional and even quasi-ballistic transport—a key requirement for integrated excitonic and quantum photonic devices.

However, achieving this level of exciton control relies critically on the quality and cleanliness of the fabricated nanostructures, particularly when implemented on fragile 2D materials such as hexagonal boron nitride (hBN)[22,23], a well-known functional substrate for van der Waals heterostructures[24], and more recently, an emerging platform for quantum photonics and sensing application[25,26]. Although advanced nanofabrication techniques such as photolithography[27], electron beam lithography (EBL)[28], ion beam lithography[29], and reactive ion etching can achieve near-10 nm resolution, they often cause structural damage and undesirable interactions with the underlying substrate during electrons or ions bombardment[29,30]. Furthermore, these methods typically involve resist layers and fluorinated etchants, which can introduce residual contamination, induce surface charging, and cause dielectric disorder-issues particularly detrimental to 2D materials. These effects are especially problematic in excitonic systems, where local dielectric inhomogeneities directly affect exciton binding energy, transport properties. While conventional laser processing offers a non-contact and maskless alternative, it has historically suffered from limited spatial resolution and rough feature boundaries due to thermal diffusion, uncontrolled ablation[31]. These drawbacks hinder its applicability in contexts requiring sharp potential profiles and minimal scattering—conditions essential for efficient exciton guidance in 2D semiconductors.

In this work, we demonstrate that femtosecond laser direct writing enables the fabrication of sub-10 nm-wide nanochannels with continuous and smooth boundaries on hBN substrates (Fig. 1b), without introducing contamination or dielectric disorder. These nanochannels modify the in-plane dielectric environment of overlaid monolayer molybdenum diselenide ($MoSe_2$), forming nanoscale exciton energy funnels that confine exciton (Fig. 1c). In a simplified physical picture, this arises because the electric field lines binding the electron–hole pair extend significantly beyond the atomic plane of the 2D material[15,32,33]. As a result, nanoscale modulation of the dielectric environment leads to spatially dependent variations in exciton binding energy, as their exciton binding energy depends on the surrounding permittivity ($\Delta E_b \propto 1/\varepsilon_{eff}^2$)[33,34]. The laser-patterned channels thus create spatial energy gradients that steer excitons along well-defined paths. Importantly, the smooth and continuous boundaries of the channels suppress electrostatic and phonon-induced scattering, thereby extending the exciton transport distance and enabling quasi-ballistic propagation over distances exceeding 5 μm (Fig. 1d).

In contrast, unconfined excitons outside the channels exhibit only short-range diffusion. Furthermore, exciton flow can be precisely redirected using custom-designed geometries, such as T-shaped junctions, demonstrating the programmability of this approach. Overall, our method provides a clean, substrate-compatible, and scalable platform for controlling exciton dynamics in 2D semiconductors and developing integrated excitonic devices.

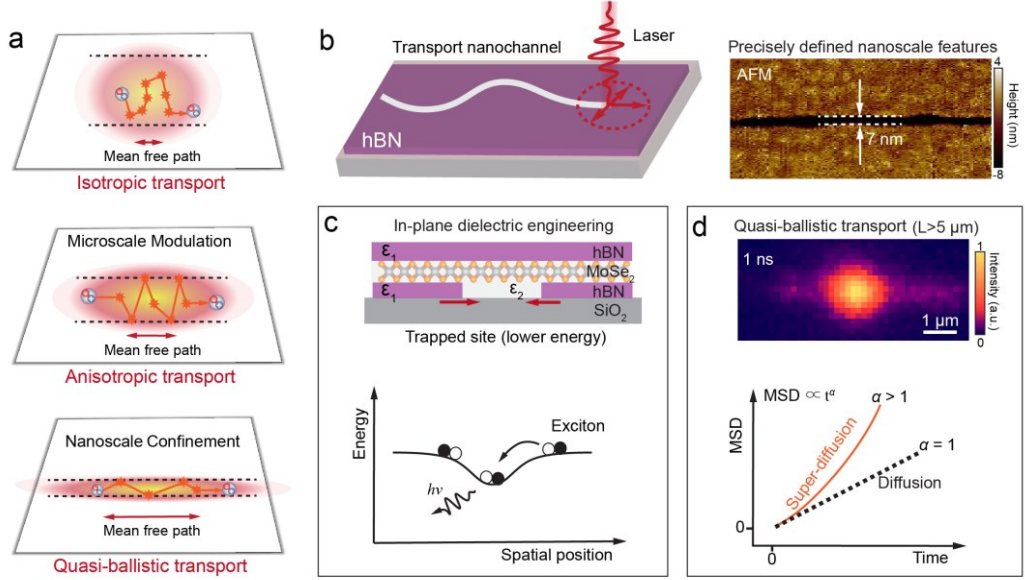

**Figure 1 | Direct-laser-written nanochannels in hBN for directional exciton transport.** (a) Conceptual illustration showing the transition from intrinsic isotropic exciton diffusion in conventional 2D semiconductors to directional quasi-ballistic transport enabled by low-energy funneling. (b) Schematic and corresponding atomic force microscopy (AFM) image of a laser-written nanochannel in hBN, showing sub-10 nm feature size, smooth boundaries, and a well-defined linear geometry. (c) Schematic illustration of in-plane dielectric engineering via laser-written nanochannel, forming low-energy trapping sites and confining excitons. (d) Exciton diffusion image acquired 1 ns after optical excitation, showing directional transport along a laser-written nanochannel embedded in hBN. Also shown are schematic comparisons of two transport regimes — random versus directional — and their corresponding mean square displacement (MSD) curves as a function of time.

## RESULTS

### Direct Laser-Processed Sub-10 nm Channels for Exciton Highways

These nanochannels, with widths of less than 10 nm, are fabricated on hBN substrates through an innovative direct laser writing technique that leverages an optical near-field coupling-induced nanostructure evolution mechanism. The laser-fabricated nanochannels are characterized by atomically sharp edges and high reproducibility, while also avoiding the charge accumulation commonly associated with electron beam or focused ion beam (FIB) techniques. The femtosecond laser processing methodology for nanochannel fabrication is illustrated in Fig. 2. In this technique, the progressive morphological evolution of femtosecond laser-induced nanostructures can be temporally controlled through pulse-to-pulse irradiation modulation. This enables programmable reconfiguration of nanochannel architectures from multiplexed arrays to isolated singular configurations, controlled by adjusting the pulse energy from 125% to 110% of the material's single-shot damage threshold ($E_{th}$)[35,36], while maintaining dimensional confinement at the sub-10 nm scale (Fig. 2a). Figure S1 presents both numerical simulations and experimental results that illustrate how the morphological changes occur gradually from one shot to the next. This phenomenon can be attributed to dynamic variations in the distribution of the electromagnetic field, resulting from the combined feedback of far-field laser irradiation and optical near-field enhancement produced by the nanostructure. As shown in Fig. 2b, the interference between the near-field scattered wave generated by the initial seed structure and the incident laser facilitates the formation of secondary seed structure along the laser polarization direction. Simultaneously, near-field enhancement drives the elongation of these structures perpendicular to the polarization axis (Fig. S2). A comprehensive discussion of the underlying mechanism is provided in the Supplementary Note 1. This effect enables the fabrication of a regular array of nanochannels with a precisely controlled number. In contrast to traditional laser processing, which typically achieves precision levels from micrometers to several hundred nanometers, obtaining sub-10-nm accuracy presents a significant challenge. The formation of nanochannels with smooth surfaces and sharp, well-defined edges is essential for controlled exciton diffusion. This is because rough edges can induce stress in two-dimensional materials, leading to exciton scattering and funneling, which ultimately impedes the directed, high-speed transport of excitons.

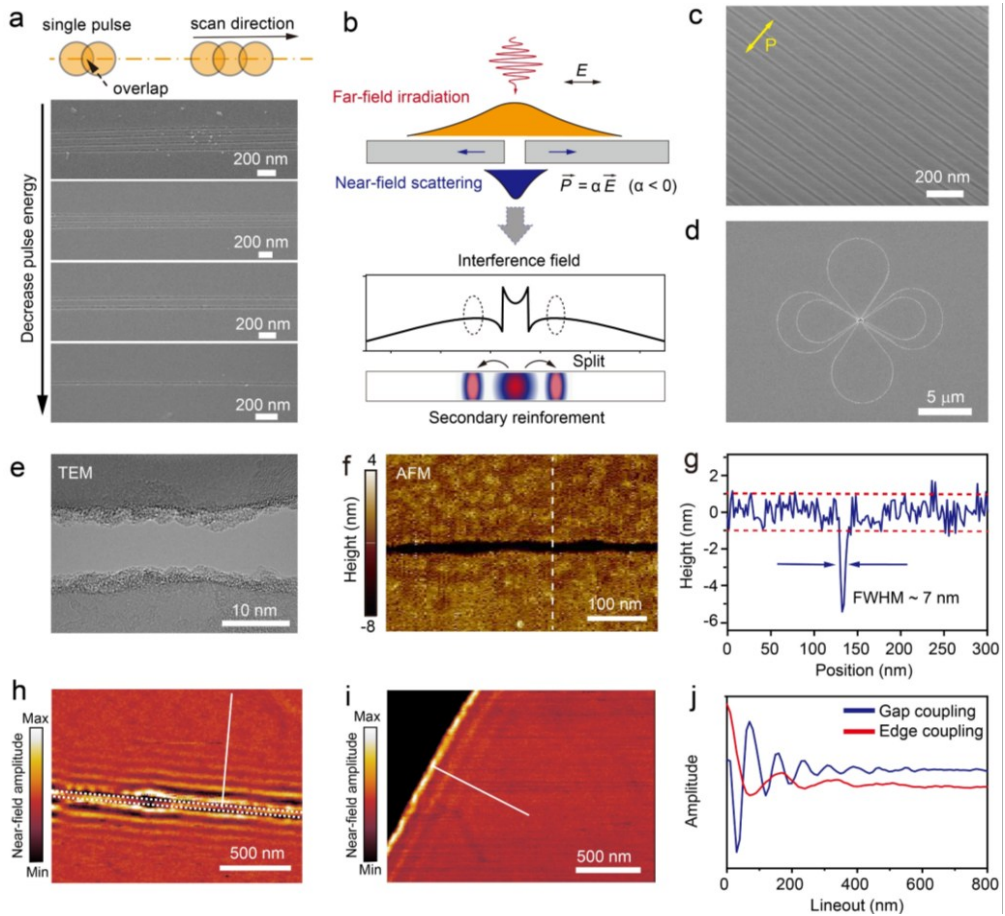

**Figure 2 | Direct laser-processed sub-10 nm free-form nanochannels with sharp edges.** (a) Laser-induced nanochannels, where the number of channels is modulated by adjusting the pulse energy (from 125% $E_{th}$ to 110% $E_{th}$). At 110% $E_{th}$, a single nanochannel is consistently formed, as the energy remains below the threshold required for secondary enhancement. (b) Schematic illustrating the interference between far-field laser irradiation and near-field scattered waves, contributing to nanochannel formation. (c) Large-area array of laser-processed nanochannels. (d) Petal-like nanostructure formed under optimized conditions. (e) TEM image of laser-induced nanochannel. (f) AFM image of a laser-fabricated nanochannel. (g) Height profile along the dashed line in (f), indicating a feature size of approximately 7 nm with atomically smooth edges. (h) Scattering-type scanning near-field optical microscopy (s-SNOM) measurements showing phonon-polariton coupling at the nanochannel edge. (i) Comparison with phonon-polariton coupling at a natural edge from the same flake. (j) Linecuts (dashed) from (h) and (i) reveal unzipped lines with atomic sharpness, exhibiting enhanced coupling to phonon-polaritons relative to natural flake edges, as evidenced by deeper modulation and slower decay.

**Free-form Nanochannels Writing with Sharp Edges**

Beyond straight-line structures, this laser writing method enables the fabrication of diverse free form nanochannels, including curved and large-scale architectures (Fig. 2c). Examples include oblique intersecting geometries and sharp-cornered configurations (Fig. S3). As previously discussed, the direction of near-field enhancement hotspots is governed by laser polarization, which is determined by the boundary conditions. Consequently, curved structures can be achieved by adjusting the laser polarization to remain perpendicular to the scanning direction (Fig. S4a). Figure S4b presents an image of closely spaced double-curved nanochannels, with separations on the order of 60 nm. Furthermore, under optimized processing conditions, a complex, intricate petal-like pattern was successfully created (Fig. 2d).

The surface roughness and the edge characteristics of the nanochannel are further characterized. High-resolution transmission electron microscopy (TEM) revealed the presence of sharp, well-defined nanoscale edges, with feature dimensions from 6 to 8 nm (Fig. 2e). Atomic force microscopy (AFM) analysis corroborates the sharpness of these edges, demonstrating atomic-level smoothness with roughness below 1 nm (Fig. 2f and Fig. 2g). To evaluate edge quality, scattering-type scanning near-field optical microscopy (s-SNOM) was employed. Figure 2h demonstrates the enhanced coupling of optical excitation to polaritons in hBN, which is attributed to the precise sharp edges of the laser-engineered nanochannels[37]. For comparison, Fig. 2i shows the coupling initiated from the natural edge of the same hBN flake. Comparative line profiles derived from s-SNOM amplitude data indicate that phonon-polaritons launched from the laser-written nanochannel exhibit significantly higher modulation depth and slower decay rates than those excited from the natural edge (Fig. 2j). Notably, laser-induced nanochannels provide superior control over edge orientation and placement compared to natural edges, presenting a versatile approach for precise polariton coupling[37]. The capability to create arrays of nanochannels with spacings on the order of tens of nanometers holds great promise for developing advanced phonon-polariton cavities and related nanophotonic devices.

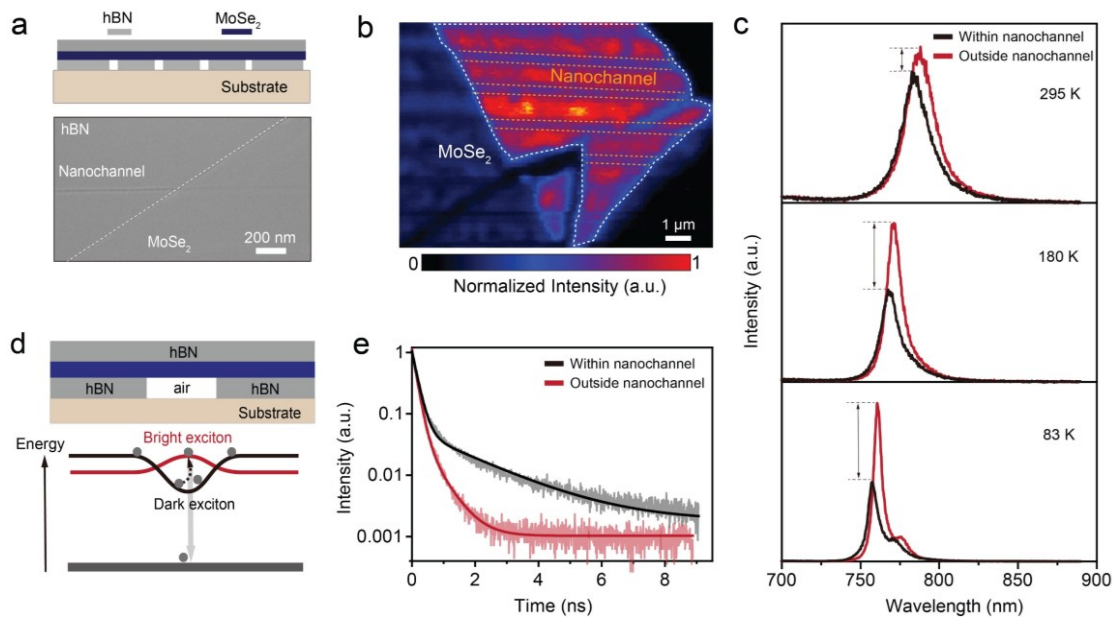

**Figure 3 | Dielectric Environment Modified Excitonic Optical Properties**. (a) Schematic illustration of the heterostructure device comprising a quartz substrate, bottom hexagonal boron nitride (hBN) layer incorporating nanostructures, monolayer MoSe$_2$, and top hBN flakes. The scanning electron microscopy (SEM) image below reveals the morphology and interface quality of the stacked layers following the transfer process. (b) Confocal photoluminescence (PL) image of the monolayer MoSe$_2$, highlighting spatial variations in emission. (c) Temperature-dependent PL spectra acquired from regions within the nanochannel and from unstructured (reference) regions, revealing distinct excitonic responses due to dielectric modulation. (d) Schematic representation of the device architecture and corresponding shifts in exciton energy levels induced by the dielectric heterostructure. (e) Time-resolved PL measurements comparing exciton lifetimes inside and outside the nanochannel, indicating modified recombination dynamics linked to the engineered dielectric environment.

**Dielectric Environment Modified Excitonic Optical Properties**

Figure 3a illustrates the sample structures used to investigate the influence of the dielectric environment on the optical properties of two-dimensional materials. A monolayer of MoSe$_2$ exfoliated from a high-quality crystal, was transferred onto an hBN bottom layer with prefabricated nanochannel structures. The SEM image presented at the bottom of Fig. 3a reveals the morphological characteristics following the transfer of the two material layers. An additional hBN layer (less than 10 nm thick) was deposited on top of the MoSe$_2$ monolayer to encapsulate the materials, which is used to reduce the effect of dielectric disorder and photo-doping on the exciton dynamics[38]. Figure 3b presents a photoluminescence (PL) image of the monolayer MoSe$_2$ measured with a custom-designed confocal microscopy setup (see Methods for details). A significant reduction in PL intensity is observed in the region above the nanochannel structure (hBN/MoSe$_2$/N-hBN). As illustrated in Fig. 3c, the PL peak observed within the nanochannel region exhibits a blue shift and reduced intensity in comparison to the region outside the nanochannel. Notably, the intensity different (ratio) between the nanochannel and surrounding region increase as the temperature decreases, which is attributed to the localization of dark exciton within the nanochannel, as discussed later. This phenomenon contrasts with the previously widely reported "strain engineering" modulation, which typically funnels the bright exciton and creates a bright site.

In our case, we attribute this to the reduced dielectric screening of the exciton in the monolayer material modified by the nanochannels. The optically excited excitons in MoSe$_2$ at room temperature are mainly neutral excitons instead of charged ones, as evidenced by temperature-dependent spectra (Fig. S5 and Supplementary Note 2). It has been demonstrated that when 2D materials like TMD are free-standing, the reduced dielectric screening can lead to strongly enhanced exciton interactions and modulate both the electronic bandgap ($\Delta E_g$) and exciton binding energy ($\Delta E_b$)[39,40]. This change in exciton behavior is because Coulomb interactions are very sensitive to the surrounding dielectric environment, as shown in previous studies[15,33,34]. In our structure, we noted that the nanochannel creates an in-plane inhomogeneity of the dielectric environment, thereby generating energy gradients that can funnel and trap excitons in nanoscale energy sites. It is important to note, however, that while dielectric engineering can influence the optical absorption and exciton binding energy of monolayer materials, the competing effects of these two factors tend to result in a relatively small net change in exciton energy (optical gap: $\Delta E_{opt} = \Delta E_g - \Delta E_b$)[33]. This is supported by the measured PL spectra of the two regions (Fig. 3c).

Considering the strong influence of the dielectric environment on dark excitons, the nanochannel region confines dark excitons, which have demonstrated efficient migration rates in prior studies[41]. In hBN encapsulated $MoSe_2$ monolayer, the spin-forbidden dark exciton lies ~2 meV above the bright one[16,42]. As discussed in the Supplementary Note 3, the change of the exciton dielectric environment within the nano-channel yields a reversal of the ordering of the bright and dark exciton states (Fig. 3d). Upon thermalization of excitons in $MoSe_2$ within the nanochannel following optical excitation, an increasing proportion will occupy the dark state within the nanochannel region. As the temperature decreases, the occupancy of dark excitons within the nanochannel becomes significantly enhanced due to their lower energy, leading to a further reduction in PL intensity[14]. This confinement of dark excitons explains the reduction of the temperature-dependence PL intensity on the nano-channel (Fig. 3c). Figure 3e presents the PL transient measurements for two regions of monolayer $MoSe_2$, within and outside the nanochannel. The data were fitted to a double-exponential decay model: $I = [A_1 \exp(-t/\tau_1)]_{fast} + [A_2 \exp(-t/\tau_2)]_{slow}$. The lifetimes for the region without the nanochannel are $\tau_1 = 0.13$ ns and $\tau_2 = 0.49$ ns, whereas the lifetimes for the region within the nanochannel were $\tau_1 = 0.16$ ns and $\tau_2 = 1.8$ ns. A ~3.7-fold increase in the slow decay process was observed in the nanochannel region. This is consistent with the expected longer lifetime of the dark excitons, which can couple to z-polarized light[43]. Notably, the luminescence from dark excitons becomes observable in this experimental configuration due to the high numerical aperture (NA) microscope objective employed. This geometry enables efficient detection of emitted light with an out-of-plane electric field component, corresponding to the normally forbidden interband optical transition associated with dark excitons (see Methods).

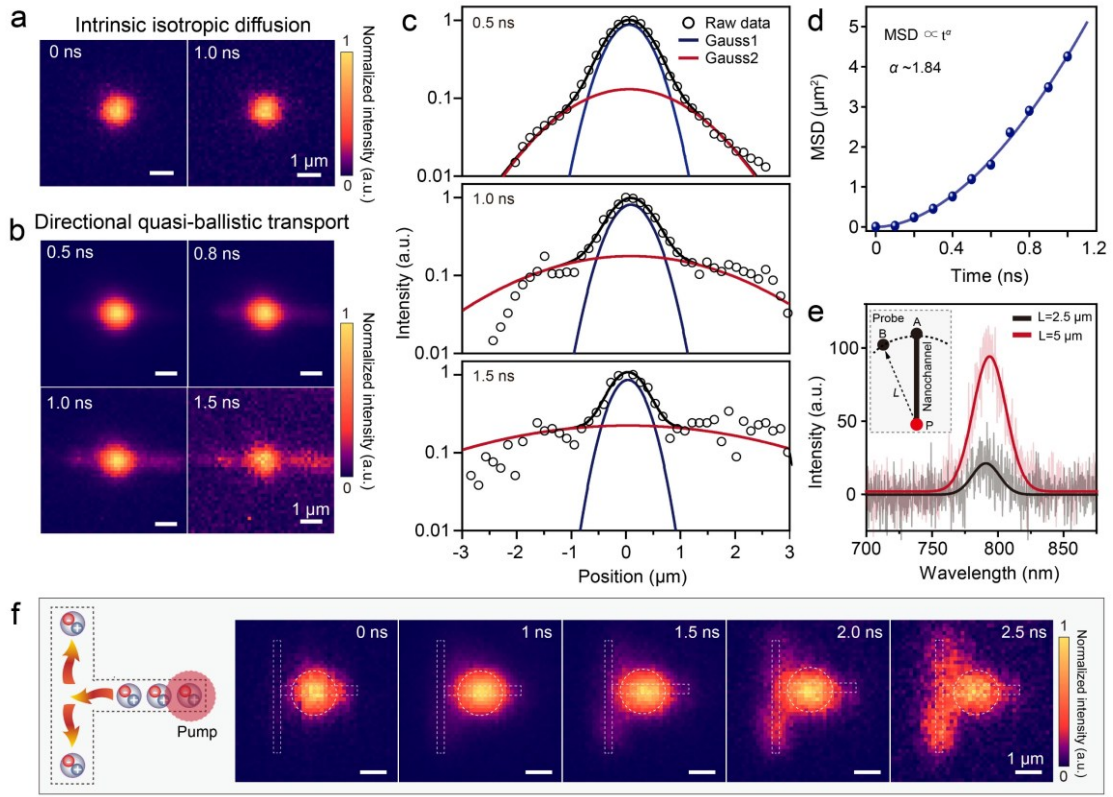

**Figure 4 | Engineered pathways for directional exciton quasi-ballistic transport**. (a) Time-resolved PL images capturing intrinsic exciton diffusion in a pristine 2D material at various time delays following pulsed optical excitation. (b) Time-resolved PL images showing directional exciton transport along an atomically defined nanochannel under identical excitation conditions. (c) Spatial profiles of exciton density extracted along the nanochannel direction from panel (b), illustrating the evolution of exciton transport over time. (d) Temporal evolution of the variance ($\sigma^2$) of the exciton distribution along the nanochannel direction, exhibiting super-diffusive behavior characterized by a power-law growth. (e) PL spectra (A–B) acquired at increasing distances from the pump location. The signal at point B is used as a spectral reference to exclude contributions from potential waveguiding effects. The inset shows the pump–probe configuration. (f) PL image of exciton transport under the influence of laser-fabricated T-shaped nanostructures, demonstrating controlled modulation and redirection of exciton propagation pathways.

**Directional Exciton Super-diffusion**

To explore the transport dynamics of the excitons regulated by the nanochannels, we employed home-built transient photoluminescence microscopy (TPLM)[44], as illustrated schematically in Figure S6. This enables direct visualization of the exciton transport process, with experimental details provided in the Methods section. To establish a reference, Fig. 4a presents the exciton diffusion behavior in an unpatterned region (hBN/MoSe$_2$/hBN), which exhibits weak and isotropic transport, consistent with prior studies[13,45]. A complete visualization of spatiotemporal dynamics is provided in Movie S1 and Fig. S7. In stark contrast, exciton transport within the nanochannel-containing region (hBN/MoSe$_2$/N-hBN) exhibits strikingly different characteristics. Spatiotemporally resolved images at various time delays post-excitation revealed pronounced, highly directional exciton propagation along the nanochannel axis (Fig. 4b; complete dynamics in Movie S2 and Fig. S8). This directional transport behavior is consistently reproducible across multiple nanochannel regions, with additional examples presented in Movie S3. The observed directional transport can be attributed to low-energy sites within the nanochannels that facilitate confined dark exciton propagation. The nanochannel architecture effectively guides excitons along its axis, demonstrating clear advantages over isotropic diffusion in pristine regions.

**Quasi-ballistic Exciton Transport along Nanochannels**

Having demonstrated that nano-channels can achieve directed transport of excitons, we further investigate the diffusivity of excitons modulated by the nanochannels. To quantify it, we fit the emission profiles at various times after excitation using double Gaussian functions, as illustrated in Fig. 4c. The time-resolved mean squared displacement (MSD), defined as MSD($t$) = $\sigma(t)^2$ - $\sigma(0)^2$, where $\sigma(t)$ represents the Gaussian width of the exciton spatial distribution at the time t after the optical excitation, is presented in Fig. 4d. In the nanochannel region, the MSD exhibits superlinear behavior, contrasting with the linear behavior characteristic of conventional diffusive transport (as shown by the diffusion without nanochannels). Fitting the MSD data to a power-law relation, MSD $\propto t^\alpha$, we find a value of $\alpha \approx 1.8$. When $\alpha = 1$, excitons undergo standard diffusive motion, characterized by a mean free path that is significantly smaller than the transport length scale. In this regime, exciton transport is affected by frequent scattering events involving defects, phonons, or other carriers[46-48]. Conversely, when $\alpha = 2$, exciton transport becomes ballistic, wherein the mean free path exceeds the characteristic transport length, resulting in highly efficient transport devoid of scattering events. In our system, the measured exponent of $\alpha \approx 1.8$ suggests quasi-ballistic transport, indicating efficient shielding from scattering and a substantial increase in the exciton mean free path within the sub-10 nm channel. Supporting this, previous studies have demonstrated that in quasi-one-dimensional channels composed of coupled quantum wells, potential fluctuations are effectively screened, thereby mitigating exciton scattering induced by impurities and defects[18]. Consequently, the exciton mean free path is extended, enhancing transport efficiency. Additionally, exciton-phonon scattering in one-dimensional potential wells has been shown to be significantly reduced, and the concomitant reduction in the phonon scattering rate further extends the exciton mean free path, reinforcing the quasi-ballistic transport regime. As a quantitative example, the diffusion length can be briefly determined from the data in the diffusion images (Fig. 4b) and the full width at half maximum (FWHM) of the fitting data (Fig. 4c) at times of 0.5 ns and 1.5 ns, yielding values of 2.05 μm and 5.03 μm, respectively. Based on these measurements, the distance traveled by the excitons during 1 ns (between 0.5 ns and 1.5 ns) can be estimated to be approximately 3 μm, assuming quasi-one-dimensional diffusion. Our data imply that the average diffusion rate during this period (1ns) is about 3000 m/s, which is about 5- 10 fold higher than the exciton drift velocities achieved by a downhill energy gradient in monolayer TMDs[41,49,50]. Thus, quasi-ballistic exciton transport achieved by these laser-induced nanochannels is highly efficient, which is important for a long diffusion length.

To quantify the exciton transport length within the device, we performed steady-state, spatially resolved photoluminescence (PL) measurements in the nanochannel region. As illustrated in Fig. 4e (Inset), excitons were excited at site P, and the corresponding PL emission from exciton recombination within the channel region (site A) was collected using a fiber-coupled spectrometer at varying distances. To eliminate potential contributions from photon transport effects, a reference signal was simultaneously acquired at an equivalent distance (site B, outside the nano-channel). The difference spectra (site A minus site B) were recorded at multiple distances under continuous-wave (CW) laser excitation with a power of 250 μW (Fig. 4e). Notably, a significant PL intensity contrast was observed at a distance of 5 μm, indicating that the exciton transport length exceeds 5 μm at room temperature. Table 1 summarizes the exciton transport properties observed in various two-dimensional materials. In typical diffusion conditions, the exciton diffusion lengths observed in 2D semiconductors are generally limited to a span of tens to a few hundred nanometers at room temperature. Only a few reports have indicated that diffusion distances of up to 1.5 μm can be observed in $WSe_2$ and $MoS_2$ monolayers [51,52] but this was measured at low temperature. Our work represents the significantly enhanced exciton diffusion length in 2D semiconductor materials, with an improvement of approximately one to two orders of magnitude compared to previous studies thanks to the quasi-one-dimensional transport. Furthermore, our study shows conclusive evidence of highly efficient and directional exciton transport in 2D materials, which is supported by the observation of quasi-ballistic transport phenomena.

**Table 1. Exciton diffusion in 2D semiconductors**

| Material | diffusion length | diffusion coefficient ($cm^2\ s^{-1}$) | refs |
|---|---|---|---|
| As-exfoliated monolayer $WSe_2$ (strained, room-temperature) | 500 nm | 0.6-1.8 | 49 |
| As-exfoliated monolayer $WSe_2$ encapsulated by hBN | 200 nm (bright) RT<br>1.5 μm (dark) low T | 14.5 (bright)<br>205 (dark) | 51 |
| As-exfoliated monolayer $WSe_2$ encapsulated by hBN | | 1.1 | 53 |
| As-exfoliated monolayer $WSe_2$ on top of a GaAs/Al0.36Ga0.64As core/shell nanowire | ~0.5 μm | 13.5 | 19 |
| As-exfoliated monolayer $WSe_2$ on a dielectric nanobubbles | 1 μm | 1.67 | 41 |
| As-exfoliated monolayer WS$_2$ on a Si/SiO$_2$ substrate | 360 nm | 0.3 | 54 |
| hBN-encapsulated WS$_2$ monolayer | | 10 | 32 |
| As-exfoliated monolayer MoSe$_2$ on SiO$_2$ | 400 nm | 12 | 13 |
| As-exfoliated monolayer MoSe$_2$ encapsulated by hBN | | 17 | 55 |
| As-exfoliated monolayer MoS$_2$ on a Si/SiO$_2$ substrate | 1.5 μm (exciton) 300 nm (Trion) | 2.1 (exciton) 18 (Trion) | 52 |
| CVD-grown monolayer MoS$_2$ on SiO$_2$/Si substrates | ~3 μm | 22.5 | 56 |
| As-exfoliated monolayer ReS$_2$ | 250 nm (Re atomic chains) 80 nm (perpendicular to Re atomic chains) | 16 (Re atomic chains) 5 (perpendicular to Re atomic chains) | 57 |
| 2D perovskite - (BA)$_2$(MA)$_{n-1}$Pb$_n$I$_{3n+1}$ | 160 nm ($n$ = 1) 670 nm ($n$ = 5) | 0.06 ($n$ = 1) 0.34 ($n$ = 5) | 58 |
| 2D perovskite - (PEA)$_2$PbI$_4$ | 236 nm | 0.192 | 59 |
| Nanoscale Dielectric-Programmed MoSe$_2$ (quasi-one-dimensional) | Above 5 μm | Super-diffusion | This work |

Note: Part of the diffusion lengths listed in the table, which are not demonstrated directly in the references, are calculated by $L = \sqrt{2D\tau}$, where $L$ is the diffusion length, $D$ is the diffusion coefficient and $\tau$ is the exciton lifetime[58].

**Programmed T-Junction for Exciton Routing in Potential Exciton Circuit**

As previously discussed, the laser can be effectively used to fabricate nanochannels with various pathways on hBN, providing a promising approach for controlling exciton routing, an essential factor for the development of excitonic devices[10,60]. To validate the practical implementation of exciton routing, we investigated the diffusion behavior within a T-junction exciton transport path. The time-resolved exciton transport image, presented in Fig. 4f, reveals the propagation dynamics of excitons along the laser-fabricated nanochannel. As anticipated, the excitons travel along the channel, and upon reaching the T-junction, bifurcate and continue propagating in two distinct directions. Comprehensive dynamic imaging data are provided in Movie S4 and Fig. S9. The successful demonstration of controlled exciton transport through precisely engineered nanochannels marks a significant step toward developing scalable technologies for optoelectronic and quantum applications.

**SUMMARY and OUTLOOK**

This work establishes a paradigm of nanoscale dielectric engineering for quantum-level control of exciton transport in 2D semiconductors, offering a platform for next-generation optoelectronic technologies. We introduce a lithography-free, near-field-enhanced laser processing technique to fabricate dielectric nanostructures with sub-10-nm precision on hBN substrates. These atomically defined dielectric nanochannels create tailored potential landscapes that significantly modify exciton dynamics in adjacent $MoSe_2$ monolayers. Notably, this dielectric confinement strategy achieves quasi-ballistic exciton transport, with diffusion lengths surpassing 5 μm - an approximately 20-fold enhancement compared to systems without nanoscale programmed dielectric channel. The direct-write fabrication technique enables the seamless integration of high-quality dielectric nanostructures with van der Waals materials, while maintaining exceptional interfacial cleanliness and structural integrity.

These technological advancements establish a powerful platform for manipulating exciton flow at the nanoscale, with significant implications for the development of advanced optoelectronic devices. The ability to engineer dielectric landscapes with nanometer precision opens exciting avenues for both fundamental research and technological applications. On the fundamental side, the interplay between atomically sharp dielectric interfaces and exciton dynamics provides a unique platform to explore quantum phenomena such as exciton superfluidity and topological exciton states. From a technological perspective, the integration of dielectric nanostructures with 2D materials holds promise for developing high-density excitonic circuits, on-chip quantum light sources, and energy-efficient photonic devices.

## MATERIALS AND METHODS

### Materials Preparation

The hexagonal boron nitride (hBN) flakes employed in this study were sourced from high-quality hBN crystals (HQ Graphene, purchased from Sixcarbon Tech, Shenzhen). These crystals were mechanically exfoliated using the Scotch-tape method to produce the desired flakes. Flakes with a thickness of approximately 20 nm were selected and transferred onto a quartz substrate for subsequent laser writing experiments aimed at nanopattern fabrication. Monolayer $MoSe_2$ material was also mechanically exfoliated using the same method and then transferred onto the patterned hBN flakes. Finally, a hBN flake with a thickness of approximately 10 nm was used to cover the entire device surface.

### Laser Writing

For the laser fabrication process, a commercial femtosecond laser (Pharos, Light Conversion) was employed as the light source. The laser was operated at a repetition rate of 1 kHz, with a single pulse extracted using a pulse picker. The laser pulses were linearly polarized along the longitudinal direction of the plane, with a pulse duration of approximately 230 fs and a wavelength of 1030 nm. The 515 nm laser converted by a second harmonic generation (SHG) process was achieved using a beta barium borate (BBO) crystal. A high numerical aperture objective lens (NA = 0.95, 50×, Olympus) was used to tightly focus the laser pulses. The full width at half maximum (FWHM) of the focal spot was measured to be approximately 350 nm. A precision translation stage enabled three-dimensional scanning. The pulse energy was controlled using a combination of a half-wave plate and a polarizer, and was monitored by a photodetector positioned prior to the objective lens. An additional half-wave plate, controlled by a rotating motor, was employed to adjust the laser polarization.

### Optical Measurement

PL and photon counting measurement: A home-built confocal microscope was utilized for optical measurements. A continuous-wave laser with a wavelength of 488 nm was employed for excitation, and the laser was focused through a high-numerical-aperture objective lens (NA = 0.95, 100×, Olympus). The PL emission was collected by the same objective lens. A 500 nm long-pass filter was inserted into the optical path to block the excitation laser. The PL signal was then directed either to a spectrometer (Princeton Instruments) for spectral analysis or to an avalanche photodiode (APD, Excelitas) coupled to a Time-Correlated Single-Photon Counting (TCSPC) module (PicoHarp 300, PicoQuantum) for photon counting. Additionally, a pulsed laser with a wavelength of 405 nm (PicoQuantum) was introduced into the system for fluorescence lifetime measurements. For temperature-dependent spectroscopic measurements (Fig. 3c), a custom-designed miniature cryogenic vacuum chamber was employed. The system was cooled using liquid nitrogen to achieve the required low-temperature conditions. A continuous-wave laser with a wavelength of 515 nm was employed for excitation, and the laser was focused by a high-numerical-aperture objective lens (NA = 0.8, 50×, Olympus). The PL signal was then directed to the spectrometer (Princeton Instruments) for the spectral analysis.

**TPLM Measurement**

In the TPLM setup, a picosecond-pulsed laser operating at 515 nm was used to optically excite excitons, focusing the beam onto the sample via an objective lens (NA = 0.65, 50×, Olympus). The emitted PL signal was collected through the same objective and directed onto the image plane. Time-resolved PL intensity at each pixel was measured using a single-photon avalanche photodiode (APD) detector integrated with a Time-Correlated Single-Photon Counting (TCSPC) module. By combining time-resolved measurements with two-dimensional scanning, we reconstructed spatially and temporally resolved images of exciton diffusion.

**AFM Measurement**

The surface morphology of the nanochannels was characterized by using atomic force microscopy (Park NX12, Park systems) in tap mode.

**TEM Measurement**

High-resolution transmission electron microscopy (TEM) was conducted on JEOL (JEM-2100F) operating at 200 kV accelerating voltage.

**SEM Measurement**

The scanning electron microscopy (SEM) images were obtained using a JEOL Emission Scanning Electron Microscope instrument (JSM-IT700HR InTouchScope). The accelerating voltage was set to 10 kV.

## ACKNOWLEDGEMENT

The authors acknowledge the financial support from the National Natural Science Foundation of China (No. 62075115, 62335013) and the National Key R&D Program of China (No. 2022YFB4600400).

## DATA AVAILABILITY

The data that support the findings of this study are available from the corresponding authors upon reasonable request.

**References**

1  Gärtner, A., Holleitner, A. W., Kotthaus, J. P. & Schuh, D. Drift mobility of long-living excitons in coupled GaAs quantum wells. *Appl. Phys. Lett.* **89**, 052108 (2006).

2  Schuetz, M. J. A., Moore, M. G. & Piermarocchi, C. Trionic optical potential for electrons in semiconductors. *Nat. Phys.* **6**, 919-923 (2010).

3  Sun, Z. *et al.* Excitonic transport driven by repulsive dipolar interaction in a van der Waals heterostructure. *Nat. Photon.* **16**, 79-85 (2022).

4  Qi, P. *et al.* Molding 2D Exciton Flux toward Room Temperature Excitonic Devices. *Adv. Mater. Technol.* **7**, 2200032 (2022).

5  Chaves, A. *et al.* Bandgap engineering of two-dimensional semiconductor materials. *npj 2D Mater. Appl.* **4**, 29 (2020).

6  Naik, M. H. & Jain, M. Substrate screening effects on the quasiparticle band gap and defect charge transition levels in $MoS_2$. *Phys. Rev. Materials* **2**, 084002 (2018).

7  Borghardt, S. *et al.* Engineering of optical and electronic band gaps in transition metal dichalcogenide monolayers through external dielectric screening. *Phys. Rev. Materials* **1**, 054001 (2017).

8  High, A. A., Hammack, A. T., Butov, L. V., Hanson, M. & Gossard, A. C. Exciton optoelectronic transistor. *Opt. Lett.* **32**, 2466-2468 (2007).

9  Yu, L. *et al.* Site-Controlled Quantum Emitters in Monolayer $MoSe_2$. *Nano Lett.* **21**, 2376-2381 (2021).

10  Pal, A. *et al.* Quantum-Engineered Devices Based on 2D Materials for Next-Generation Information Processing and Storage. *Adv. Mater.* **35**, 2109894 (2023).

11  Mueller, T. & Malic, E. Exciton physics and device application of two-dimensional transition metal dichalcogenide semiconductors. *npj 2D Mater. Appl.* **2**, 29 (2018).

12  Xia, F., Wang, H., Xiao, D., Dubey, M. & Ramasubramaniam, A. Two-dimensional material nanophotonics. *Nat. Photon.* **8**, 899-907 (2014).

13  Kumar, N. *et al.* Exciton diffusion in monolayer and bulk $MoSe_2$. *Nanoscale* **6**, 4915-4919

(2014).

14. Malic, E. *et al.* Dark excitons in transition metal dichalcogenides. *Phys. Rev. Materials* **2**, 014002 (2018).

15. Raja, A. *et al.* Dielectric disorder in two-dimensional materials. *Nat. Nanotechnol* **14**, 832-837 (2019).

16. Lu, Z. *et al.* Magnetic field mixing and splitting of bright and dark excitons in monolayer $MoSe_2$. *2D Mater.* **7**, 015017 (2020).

17. Kato, T. & Kaneko, T. Transport Dynamics of Neutral Excitons and Trions in Monolayer $WS_2$. *ACS Nano* **10**, 9687-9694 (2016).

18. Vögele, X. P., Schuh, D., Wegscheider, W., Kotthaus, J. P. & Holleitner, A. W. Density Enhanced Diffusion of Dipolar Excitons within a One-Dimensional Channel. *Phys. Rev. Lett.* **103**, 126402 (2009).

19. Dirnberger, F. *et al.* Quasi-1D exciton channels in strain-engineered 2D materials. *Sci. Adv.* **7**, eabj3066 (2021).

20. Dai, Y. *et al.* Two-Dimensional Exciton Oriented Diffusion via Periodic Potentials. *ACS Nano* **18**, 23196-23204 (2024).

21. Wurdack, M. *et al.* Motional narrowing, ballistic transport, and trapping of room-temperature exciton polaritons in an atomically-thin semiconductor. *Nat. Commun.* **12**, 5366 (2021).

22. Caldwell, J. D. *et al.* Photonics with hexagonal boron nitride. *Nature Reviews Materials* **4**, 552-567 (2019).

23. Sortino, L. *et al.* Atomic-layer assembly of ultrathin optical cavities in van der Waals heterostructure metasurfaces. *Nat. Photon.* (2025).

24. Dean, C. R. *et al.* Boron nitride substrates for high-quality graphene electronics. *Nat. Nanotechnol* **5**, 722-726 (2010).

25. Tran, T. T., Bray, K., Ford, M. J., Toth, M. & Aharonovich, I. Quantum emission from hexagonal boron nitride monolayers. *Nat. Nanotechnol* **11**, 37-41 (2016).

26. Fang, H.-H., Wang, X.-J., Marie, X. & Sun, H.-B. Quantum sensing with optically accessible spin defects in van der Waals layered materials. *Light: Science & Applications* **13**, 303 (2024).

27. Dathbun, A. *et al.* Selectively Metallized 2D Materials for Simple Logic Devices. *ACS Applied Materials & Interfaces* **11**, 18571-18579 (2019).

28. Fröch, J. E., Hwang, Y., Kim, S., Aharonovich, I. & Toth, M. Photonic Nanostructures from Hexagonal Boron Nitride. *Adv. Opt. Mater.* **7**, 1801344 (2019).

29. Glushkov, E. *et al.* Engineering Optically Active Defects in Hexagonal Boron Nitride Using Focused Ion Beam and Water. *ACS Nano* **16**, 3695-3703 (2022).

30. Meyer, J. C. *et al.* Accurate Measurement of Electron Beam Induced Displacement Cross Sections for Single-Layer Graphene. *Phys. Rev. Lett.* **108**, 196102 (2012).

31. Malinauskas, M. *et al.* Ultrafast laser processing of materials: from science to industry. *Light: Science & Applications* **5**, e16133-e16133 (2016).

32. Zipfel, J. *et al.* Exciton diffusion in monolayer semiconductors with suppressed disorder. *Phys. Rev. B* **101**, 115430 (2020).

33. Cho, Y. & Berkelbach, T. C. Environmentally sensitive theory of electronic and optical transitions in atomically thin semiconductors. *Phys. Rev. B* **97**, 041409 (2018).

34. Raja, A. *et al.* Coulomb engineering of the bandgap and excitons in two-dimensional materials. *Nat. Commun.* **8**, 15251 (2017).

35  Wang, X.-J., Fang, H.-H., Li, Z.-Z., Wang, D. & Sun, H.-B. Laser manufacturing of spatial resolution approaching quantum limit. *Light: Science & Applications* **13**, 6 (2024).

36  Li, Z.-Z. *et al.* O-FIB: far-field-induced near-field breakdown for direct nanowriting in an atmospheric environment. *Light: Science & Applications* **9**, 41 (2020).

37  Chen, C. Y. *et al.* Unzipping hBN with ultrashort mid-infrared pulses. *Sci. Adv.* **10**, eadi3653 (2024).

38  Cadiz, F. *et al.* Excitonic Linewidth Approaching the Homogeneous Limit in $MoS_2$-Based van der Waals Heterostructures. *Phys. Rev. X* **7**, 021026 (2017).

39  Gerber, I. C. & Marie, X. Dependence of band structure and exciton properties of encapsulated $WSe_2$ monolayers on the hBN-layer thickness. *Phys. Rev. B* **98**, 245126 (2018).

40  Sun, X. *et al.* Enhanced interactions of interlayer excitons in free-standing heterobilayers. *Nature* **610**, 478-484 (2022).

41  Su, H. *et al.* Dark-Exciton Driven Energy Funneling into Dielectric Inhomogeneities in Two-Dimensional Semiconductors. *Nano Lett.* **22**, 2843-2850 (2022).

42  Robert, C. *et al.* Measurement of the spin-forbidden dark excitons in $MoS_2$ and $MoSe_2$ monolayers. *Nat. Commun.* **11**, 4037 (2020).

43  Wang, G. *et al.* In-Plane Propagation of Light in Transition Metal Dichalcogenide Monolayers: Optical Selection Rules. *Phys. Rev. Lett.* **119**, 047401 (2017).

44  Li, X.-Z. *et al.* Optical Visualization of Photoexcitation Diffusion in All-Inorganic Perovskite at High Temperature. *J. Phys. Chem. Lett.* **13**, 7645-7652 (2022).

45  Wagner, K. *et al.* Nonclassical Exciton Diffusion in Monolayer $WSe_2$. *Phys. Rev. Lett.* **127**, 076801 (2021).

46  Ginsberg, N. S. & Tisdale, W. A. Spatially Resolved Photogenerated Exciton and Charge Transport in Emerging Semiconductors. *Annu. Rev. Phys. Chem.* **71**, 1-30 (2020).

47  Mantsevich, V. N. & Glazov, M. M. Viscous hydrodynamics of excitons in van der Waals heterostructures. *Phys. Rev. B* **110**, 165305 (2024).

48  del Águila, A. G. *et al.* Ultrafast exciton fluid flow in an atomically thin $MoS_2$ semiconductor. *Nat. Nanotechnol* **18**, 1012-1019 (2023).

49  Cordovilla Leon, D. F., Li, Z., Jang, S. W., Cheng, C.-H. & Deotare, P. B. Exciton transport in strained monolayer $WSe_2$. *Appl. Phys. Lett.* **113**, 252101 (2018).

50  Moon, H. *et al.* Dynamic Exciton Funneling by Local Strain Control in a Monolayer Semiconductor. *Nano Lett.* **20**, 6791-6797 (2020).

51  Cadiz, F. *et al.* Exciton diffusion in $WSe_2$ monolayers embedded in a van der Waals heterostructure. *Appl. Phys. Lett.* **112**, 152106 (2018).

52  Uddin, S. Z. *et al.* Neutral Exciton Diffusion in Monolayer $MoS_2$. *ACS Nano* **14**, 13433-13440 (2020).

53  Wagner, K. *et al.* Diffusion of Excitons in a Two-Dimensional Fermi Sea of Free Charges. *Nano Lett.* **23**, 4708-4715 (2023).

54  Kulig, M. *et al.* Exciton Diffusion and Halo Effects in Monolayer Semiconductors. *Phys. Rev. Lett.* **120**, 207401 (2018).

55  Hotta, T. *et al.* Exciton diffusion in hBN-encapsulated monolayer $MoSe_2$. *Phys. Rev. B* **102**, 115424 (2020).

56  Yu, Y. *et al.* Giant enhancement of exciton diffusivity in two-dimensional semiconductors. *Sci. Adv.* **6**, eabb4823 (2020).

57  Cui, Q. *et al.* Transient Absorption Measurements on Anisotropic Monolayer ReS$_2$. *Small* **11**, 5565-5571 (2015).

58  Deng, S. *et al.* Long-range exciton transport and slow annihilation in two-dimensional hybrid perovskites. *Nat. Commun.* **11**, 664 (2020).

59  Seitz, M. *et al.* Exciton diffusion in two-dimensional metal-halide perovskites. *Nat. Commun.* **11**, 2035 (2020).

60  Lamsaadi, H. *et al.* Exciton Collimation, Focusing and Trapping Using Complex Transition Metal Dichalcogenide Lateral Heterojunctions. *Adv. Opt. Mater.* **13**, 2403009 (2025).